\documentclass[twocolumn,prl,showpacs,amsmath]{revtex4}
\usepackage{epsfig,amsfonts,amsbsy}
\newcommand{\bra}{\langle}
\newcommand{\ket}{\rangle}
\newcommand{\pder}[2]{\frac{\partial #1}{\partial  #2}}
\newcommand{\pderf}[3]{\left(\frac{\partial #1}{\partial  #2}\right)_{#3}}

\newcommand{\bv}[1]{{\boldsymbol #1}}
\newcommand{\e}{{\rm e}}

\newcommand{\ep}{\epsilon}
\newcommand{\cS}{{\cal S}}

\begin{document}
\title{Derivation of hydrodynamics from the Hamiltonian description 
of particle systems}
\author{Shin-ichi Sasa}
\affiliation {
Department of Physics, Kyoto University, Kyoto 606-8502, Japan}
\date{\today}

\begin{abstract}
Hamiltonian particle systems may exhibit non-linear 
hydrodynamic phenomena as the time evolution of the 
density fields of energy, momentum, and mass.
In this Letter, an exact equation describing the 
time evolution is derived assuming the local Gibbs 
distribution at initial time. The key concept in the 
derivation is an identity 
similar to the fluctuation theorems. The Navier-Stokes equation 
is obtained as a result of  simple perturbation expansions 
in a small parameter that represents the scale separation.
\end{abstract}

\maketitle

\paragraph{Introduction:}


Let us consider a large number of interacting  particles 
that obey a classical Hamiltonian equation as an isolated 
system. The initial  condition is assumed to provide spatial 
variation in the macroscopic density fields. Examples include 
turbulent configurations of the momentum density field. Through 
time evolution of the particles, the density fields exhibit interesting 
space-time structures, and  the system eventually relaxes to 
an equilibrium state. This process is believed to be described 
universally by hydrodynamic equations of the density fields 
for  liquids and gases \cite{Landau}. As a stimulating 
numerical experiment of a Hamiltonian particle system,
quite recently, the Kolmogorov spectrum has been observed in 
a transient state \cite{Komatsu}. 
However, it remains unclear whether the hydrodynamic 
description of such violent time evolution can be understood  
on the basis of Hamiltonian particle systems.


The problem of deriving the hydrodynamic equations 
has been studied over the last century by various
approaches such as the analysis of the Boltzmann equation \cite{Grad},
the calculation on the basis of local equilibrium
distributions \cite{Kirkwood,Morray},
and the derivation from hypothetical non-equilibrium ensembles 
\cite{Mori, Mclennan,Zubarev,KG}. However, 
non-linear hydrodynamic phenomena of liquids are out of scope of 
these classical works, because the Boltzmann equation  applies  
to only dilute gases, the local equilibrium distributions cannot 
describe dissipation effects, and the hypothetical ensembles are 
not justified  beyond the linear response regime from 
a global equilibrium state \cite{Maes2}.


Nevertheless, a physical intuition of the hydrodynamic 
description in particle systems is simple, and it has
been known as follows. When the length scale of the spatial 
variation  of the density fields is much larger than the particle scales, 
the particle distribution may be characterized by a local 
equilibrium distribution with local thermodynamic variables. 
If such local equilibrium propagates in time, the solution 
of the Liouville equation may be close to the local equilibrium 
distribution \cite{Spohn-book}. For example, turbulence 
is a strongly non-linear phenomenon described by a deterministic 
equation for the density fields, 
whereas the particle distribution in the turbulent state is still 
near local equilibrium. Thus, the non-linear hydrodynamic 
equations may be derived  by developing a perturbation theory
with a small parameter representing the scale separation. 


As one attempt that follows the physical intuition,
a perturbation  theory leading to the incompressible 
Navier-Stokes equation was presented with the assumption
of  the propagation of local equilibrium \cite{Esposito}. 
Since this study  attempts  to construct solutions of the 
Liouville equation  without specifying the initial condition, 
it seems difficult to investigate the validity of the assumption. 
Rather, in order to obtain a definite result, 
it may be necessary that a special class of initial conditions
is focused on.


Such a situation can be seen in the statistical 
mechanical derivation of the second law of thermodynamics.
If an initial distribution satisfies some conditions,
the second law of thermodynamics can be derived \cite{Lenard,Tasaki}. 
In particular, if the canonical distribution is assumed initially, 
the proof of the second law becomes  quite simple by utilizing
the Jarzynski equality \cite{Jarzynski}. 
These results obtained for special but familiar initial
distributions provide key insights into the second law 
of thermodynamics.


In this Letter, as an extension of these studies,
the nonlinear hydrodynamic equations are derived in a quite 
compact manner for an isolated Hamiltonian system 
(i.e., a purely mechanical system) with an initial 
condition given by the local Gibbs distribution.  
The crucial step in the derivation is to find a universal 
relation  similar to the fluctuation theorems 
\cite{Evans,Gallavotti,Kurchan,Maes,LS,Crooks,d-FT,Seifert}, 
by which an exact expression for the time evolution of 
the density fields is obtained. Furthermore, 
a straightforward perturbation leads 
to the Navier-Stokes equation after
introducing a scale separation  parameter 
that is assumed to be small.


\paragraph{Framework:}


Let $\bv{r}_i$ and $\bv{p}_i$ $(1 \le i \le N)$ be the position and 
the momentum of $i$-th particle with the mass $m$.  
The short-range interaction potential between $i$-th and $j$-th 
particles is expressed as $V(|\bv{r}_{ij}|)$, 
where $\bv{r}_{ij}=\bv{r}_i-\bv{r}_j$. The phase space coordinate
of the system is a collection of $(\bv{r}_i, \bv{p}_i)$,
which is denoted by $\Gamma$. The particles are confined in 
a cube $\Omega$ with length $L$. For simplicity, periodic
boundary conditions are assumed. The time evolution of
the system is described by the Hamiltonian equation. The
solution of the equation for an initial state $\Gamma$ 
is simply denoted as $\Gamma_t$ for any real number $t$.  
That is, $\Gamma_{t+s}= (\Gamma_{t})_{s}$ for any $t$ and $s$.


There are  five conserved quantities: 
the total mass, the total momentum, and  the total energy.
It is assumed that there are no other independent 
conserved quantities and that there are no soft modes
arising from symmetry breaking. 
Corresponding to the conserved quantities, 
the following microscopic density 
fields are defined: $\hat \rho(\bv{r};\Gamma)
\equiv  \sum_{i} m \delta(\bv{r}-\bv{r}_i)$,
$\hat {\bv{\pi}}(\bv{r};\Gamma)
\equiv \sum_{i} \bv{p}_i \delta(\bv{r}-\bv{r}_i)$, 
and 
\begin{equation}
\hat h(\bv{r};\Gamma)
\equiv \sum_{i} 
\left[\frac{p_i^2}{2m} +\frac{1}{2}\sum_{j \not =i} V(|\bv{r}_{ij}|) \right]
\delta(\bv{r}-\bv{r}_i).
\end{equation}
A collection of the five density fields is denoted  by 
$\hat C = (\hat h, \hat{\bv{\pi}}, \hat \rho)$. 
A component of $\hat C$ is expressed as $\hat C^\alpha$, 
where  $\alpha \in \{0, 1,2,3,4\}$.
The density fields  
$\hat C^\alpha(\bv{r};\Gamma)$ satisfy 
the continuity equation
\begin{equation}
\partial_t \hat C^\alpha(\bv{r};\Gamma_t)+
\partial^a \hat J^{\alpha a}(\bv{r};\Gamma_t)=0,
\end{equation}
where the microscopic 
currents $\hat J^{\alpha a }(\bv{r};\Gamma)$ can be written
explicitly in terms of $\Gamma$ \cite{supp}. 
The roman alphabet $a$, $b$, and $c$
take values $1$, $2$, or $3$, which represents the index of the Euclidian 
coordinates. 
Here and hereafter, the same index appearing in one term
indicates that the summation with respect to this index 
is taken. For the sake of notation simplicity, 
$f\cdot g \equiv \int_{\Omega} d^3\bv{r} f(\bv{r}) g(\bv{r})$. 


The main assumption in this Letter is that 
initial states $\Gamma$ at $t=0$ are chosen 
according to the local Gibbs distribution
\begin{equation}
P_{\rm LG}(\Gamma;\lambda) 
=\e^{-\lambda^\alpha \cdot \hat C^\alpha(\Gamma)-\Psi(\lambda)},
\label{initial}
\end{equation}
where the position dependent parameter  $\lambda(\bv{r})$ 
characterizes the inhomogeneous distribution of the density fields.
The normalization condition of the probability determines
the functional $\Psi(\lambda)$ for the Hamiltonian. Let us call 
$\lambda$ and $\Psi(\lambda)$ the conjugate field and 
the Mathieu functional, respectively, although these 
names may not be accurate except for the limited case where 
$\lambda$ is spatially homogeneous. $\Psi(\lambda)$ is assumed 
to be a convex  functional of $\lambda$ as a natural
extension from the spatially homogeneous case.
One example of initial states may be a snap shot of 
molecular configuration in turbulent state.

The local Gibbs distribution describes an inhomogeneous equilibrium 
state in the moving frame with the local velocity 
$\bv{u}\equiv -\bv{\lambda}/\lambda^0 $. 
Indeed,  let $\Gamma'$ be the phase space coordinate obtained from 
$\Gamma$ by transforming 
$\bv{p}_i$ in $\Gamma$ to $\bv{p}_i -m \bv{u}(\bv{r}_i)$. 
In general, a dynamical variable in the moving frame is defined as 
$\hat A'(\Gamma) \equiv \hat A(\Gamma')$ for any dynamical variable
$\hat A$. Then, $\lambda^\alpha\cdot \hat C^\alpha$
is written  as 
\begin{equation}
\lambda^\alpha\cdot \hat C^\alpha=
\lambda^0 \cdot \hat h' + 
\left( \lambda^4-\frac{1}{2} 
\frac{|\bv{\lambda}|^2}{\lambda^0} \right)
\cdot \hat \rho,
\label{lambda-C}
\end{equation}
which implies that $\lambda^0(\bv{r})$ and $\lambda^4(\bv{r})$
are related to inhomogeneous inverse temperature and a one-body 
potential function. It should be noted that $\hat h'$ is  
interpreted as the internal energy density field. 


The expectation value of the density field $\hat C(\bv{r};\Gamma_t)$
at time $t$ with respect to the initial distribution is 
given by  $ \int d\Gamma  P_{\rm LG}(\Gamma;\lambda) \hat C(\bv{r};\Gamma_t)$.
It is further  written as 
$ \int d\Gamma  P_{t}(\Gamma) \hat C(\bv{r};\Gamma)$, where 
\begin{equation}
P_t(\Gamma)=
\e^{-\lambda^\alpha \cdot \hat C^\alpha(\Gamma_{-t})-\Psi(\lambda)}
\label{pt}
\end{equation}
using the Liouville theorem $|d\Gamma_t/d\Gamma|=1$.  
The expectation value with respect to $P_{\rm LG}(\Gamma;\lambda)$
and $P_t(\Gamma)$ are denoted by $\bra \  \ket_{\lambda}^{\rm LG}$ 
and $\bra \ \ket_t$, respectively.
Then, the time evolution of 
$C^\alpha_t (\bv{r}) $, which represents 
$ \bra \hat C^\alpha (\bv{r}) \ket_t$, is derived as
\begin{equation}
\partial_t C^\alpha_t(\bv{r})+
\partial^a J^{\alpha a}_t(\bv{r})=0,
\label{exp-con}
\end{equation}
where $J^{\alpha a}_t (\bv{r})$ is found to be equal to 
$\bra \hat J^{\alpha a} (\bv{r}) \ket_t$.
Throughout this Letter, the dynamical variable 
with the hat symbol, e.g., $\hat A$, indicates 
that this variable depends on the phase space coordinate $\Gamma$,
whereas the quantity without the hat symbol indicates an expectation value.
When $J_t^{\alpha a}$ is expressed in terms of $C_t$,
(\ref{exp-con}) becomes a hydrodynamic equation.

\paragraph{Main result:}

I  present three identities in the framework.
The first identity  is 
\begin{equation}
\partial^a \lambda^\alpha \cdot \bra \hat J^{\alpha a} \ket_{\lambda}^{\rm LG}=0
\label{reversible}
\end{equation}
for any $\lambda$, which can be derived from 
the trivial relation
$\partial_t \left[ \left. \int d\Gamma P_t(\Gamma) \right] \right|_{t=0}=0$.


The second identity originates from  statistical mechanical
formulas of the local Gibbs distribution.  In order to 
utilize the formulas,  the time evolution of the density fields
is represented  by using the time dependent conjugate fields 
$\lambda_t^\alpha$, which is defined by
$C_t^\alpha(\bv{r})= \bra \hat C^\alpha(\bv{r}) \ket_{\lambda_t}^{\rm LG}$.
This definition is then written as
\begin{equation}
C^\alpha_t(\bv{r})
= - 
\left.
\frac{\delta \Psi(\lambda)}{\delta \lambda^\alpha(\bv{r})}
\right\vert_{\lambda=\lambda_t} .
\label{ext-con}
\end{equation}
Owing to the convexity of the Mathieu functional
$\Psi(\lambda)$, (\ref{ext-con}) is
further expressed as a variational equation:
\begin{equation}
\lambda_t = {\rm Arginf}_\lambda 
[ \lambda^\alpha \cdot C_t^\alpha + \Psi(\lambda) ].
\label{sigmainf}
\end{equation}
This  naturally leads to the definition of the functional
\begin{equation}
\cS(C) \equiv \inf_\lambda [ \lambda^\alpha \cdot C^\alpha + \Psi(\lambda) ].
\label{S-def}
\end{equation}
When the density field $C$ is spatially homogeneous, 
$\cS(C)$ is identical to the thermodynamic entropy. 
In this sense, $\cS$ may be an extension of the  entropy
to that for non-equilibrium states characterized by 
the density field, but  no thermodynamic interpretation is given yet.
The important thing here is that the quantity $\cS(C)$ 
is determined from the Hamiltonian under consideration. 
Then, for  a given density field $C_t(\bv{r})$, 
the conjugate field $\lambda_t (\bv{r})$ is calculated by
\begin{equation}
\lambda_t (\bv{r}) 
=  \left. \frac{\delta \cS(C)}{\delta C(\bv{r})} \right|_{C=C_t},
\label{lambda-def}
\end{equation}
which is the second identity.
At this stage, let us remind that $C_t$ and $\lambda_t$
are determined uniquely from the time evolution and 
the initial distribution. 

The third identity is a variant of the so-called 
fluctuation theorems. In order to obtain it,
(\ref{pt})  is formally rewritten  as
\begin{equation}
P_t(\Gamma)
=\e^{-\lambda_t^\alpha \cdot {\hat C}^\alpha(\Gamma)-\Psi(\lambda_t)+\hat \Sigma_t(\Gamma)},
\label{pt-2}
\end{equation}
where 
\begin{equation}
\hat \Sigma_t(\Gamma)=\int_0^t ds 
\partial_s[\lambda_s^\alpha \cdot 
\hat C^\alpha (\Gamma_{s-t})+\Psi(\lambda_s)].
\label{Sigma-def} 
\end{equation}
By comparing  (\ref{pt-2}) with (\ref{pt}), 
it is found for any dynamical variable $\hat A (\Gamma)$
that 
\begin{equation}
\bra \hat A \ket_t =\bra \hat A e^{\hat \Sigma_t} \ket_{\lambda_t}^{\rm LG},
\label{a-ave}
\end{equation}
which is the third identity. Various non-trivial relations 
of statistical quantities can be derived from this identity.
As one example, 
(\ref{a-ave}) with $\hat A=\e^{-\hat \Sigma_t}$
becomes $\bra e^{- \hat \Sigma_t} \ket_t  =1$, which corresponds 
to an integral fluctuation theorem. With a  trivial inequality
$e^{-x} \ge 1-x$, the integral fluctuation theorem leads to the 
inequality $ \bra \hat \Sigma_t \ket_t \ge 0$. Since 
(\ref{Sigma-def}) gives $\bra \Sigma_t \ket_t=
\cS(C_t) - \cS(C_0)$,  $ \bra \hat \Sigma_t \ket_t \ge 0$ implies 
$\cS(C_t) \ge \cS(C_0) $
for any $t$. This can  be related to the second law
of thermodynamics for some special cases.


These three identities provide a useful representation
of the time evolution. Given $C_s$ for $s \in [0,t]$,
one determines $\lambda_s$ by the second identity 
(\ref{lambda-def}). Then, the third identity  (\ref{a-ave}) 
leads to 
\begin{equation}
J^{\alpha a}_t (\bv{r})= \bra  \hat J^{\alpha a} (\bv{r})
\e^{\hat \Sigma_t}\ket^{\rm LG}_{\lambda_t}
\label{NNRF}
\end{equation}
with 
\begin{equation}
\hat \Sigma_t(\Gamma)
= \int_0^t ds 
\left[ 
(\partial^a \lambda_s^{\alpha})\cdot \delta \hat J^{\alpha a}_s
+(\partial_s \lambda_s^\alpha) \cdot \delta \hat C^\alpha_s 
\right],
\label{Sigma-def2} 
\end{equation}
where $\delta \hat A_s
\equiv \hat A(\Gamma_{s-t})- \bra \hat A \ket_{\lambda_s}^{\rm LG}$
for any $\hat A$, and the first identity (\ref{reversible}) and 
the formula (\ref{ext-con})  have been used in the 
derivation of (\ref{Sigma-def2}). 
The continuity equation (\ref{exp-con}) leads to 
$\partial_t C^\alpha_t(\bv{r})$ from $J^{\alpha a}_t(\bv{r})$. 
Expressions similar to the non-equilibrium 
ensemble (\ref{pt-2}) with (\ref{Sigma-def2}) have been proposed
in many studies such as Refs. \cite{Mclennan,Zubarev,KG}.
(See Ref. \cite{Maes2} for the rigorous characterization on 
the so-called Mclennan ensemble.)
However, in this Letter, purely Hamiltonian systems are 
considered without interaction with external reservoirs, 
in contrast to previous studies.


It should be stressed that the general result 
consisting of (\ref{exp-con}), (\ref{lambda-def}),
(\ref{NNRF}), and (\ref{Sigma-def2})
can be obtained without any complicated calculation.  
This {\it exact} evolution equation of $C_t$ will be a 
starting point for the analysis of all hydrodynamic 
phenomena which may  include non-standard cases
that are not described by the Navier-Stokes equation.
Moreover, this expression immediately leads to the 
well-known hydrodynamic equations when a small parameter 
representing the scale separation is introduced, 
which is explained below.


\paragraph{Perturbation theory:}


Let $\xi_{\rm micro}$ be the maximum length scale appearing
in the molecular description such as 
the molecule size, the interaction length, or the mean free path,
and  $\xi_{\rm macro}$ be the minimum length characterizing  
macroscopic behaviors. A scale separation parameter is 
then introduced as  $\ep \equiv \xi_{\rm micro}/\xi_{\rm macro}$,
which is assumed to be small. 
Hereafter, all the quantities are assumed to be 
dimensionless  by setting $\xi_{\rm micro}=1$, $m=1$, and 
$\lambda^0(\bv{r}_0)=1$ for some $\bv{r}_0 \in \Omega$.


Macroscopic density fields are initially generated by choosing 
$\Gamma$ according to the local Gibbs distribution 
with the choice $\lambda^\alpha(\bv{r})=\bar \lambda^\alpha(\ep \bv{r})$,
where the functional form of $\bar \lambda^\alpha$ is 
independent of $\epsilon$. All the small quantities originate
from this $\epsilon$. For example, $\partial^a C_s^\alpha \simeq O(\ep)$
and $\partial^a \lambda_s^\alpha \simeq O(\ep)$ are reasonably
conjectured for any $s$. Based on these estimations, 
any quantity $A$ can be expanded as 
$A=A^{(0)}+A^{(1)}+A^{(2)}+O(\ep^3)$,
where $A^{(k)}/\ep^k$  is finite for $\ep \to 0$.


First, the functional $\cS$ is expanded. 
Since  $\cS$ may be expressed
as a space integration of some functions of the density fields 
and their spatial derivatives,  
$\cS^{(1)}=0$ from the reflection symmetry,
$\cS^{(2)}$ contains terms such as $(\partial^a C^\alpha)^2$,
and  $\cS^{(0)}$ consists of terms without  spatial
derivatives. That is, $\cS^{(0)}$ is equal to  the space integration
of the local entropy density,
where  (\ref{lambda-C})  and (\ref{S-def}) should be
noticed. Explicitly, 
by using the thermodynamic entropy density $s_{\rm th}$
as a function of $h'$ and $\rho$ for the system,
$\cS^{(0)}$ is expressed as
\begin{equation}
\cS^{(0)}(C) = \int_\Omega d^3 \bv{r} s_{\rm th}(h'(\bv{r}), \rho(\bv{r})).
\label{cs0}
\end{equation}
According to thermodynamics, the inverse temperature and the 
chemical potential are defined by 
$\beta \equiv\partial  s_{\rm th}(h', \rho)/\partial h' $ 
and $ \mu \equiv -\beta^{-1} \partial s_{\rm th}(h', \rho)/\partial\rho$,
respectively.  Through this relation, $\beta_s(\bv{r})$ and $\mu_s(\bv{r})$
are determined from $C_s^\alpha(\bv{r})$. 
Then, from (\ref{lambda-def}) and (\ref{cs0}) ,
$\beta_s$ and $\mu_s$ are related to the 
conjugate fields as 
$\beta_s=\lambda_s^0 +O(\ep^2)$ and 
$\beta_s \mu_s =-\lambda_s^4+ \beta_s |\bv{u}_s|^2/2+O(\ep^2)$,  
and the straightforward calculation \cite{supp} using a thermodynamic 
relation yields 
\begin{eqnarray}
& &(\partial^a \lambda_s^\alpha) \cdot J_s^{\alpha a}
\nonumber \\
&=& 
(\partial^a\beta_s) \cdot J_s^{0a}{}' 
-(\beta_s\partial^b u_s^a) \cdot(J_s^{ab}{}' -p_s\delta^{ab})+O(\ep^3),
\label{sigma-def}
\end{eqnarray}
where $p(\bv{r})$ is the thermodynamic pressure determined from
$s_{\rm th}(h'(\bv{r}), \rho(\bv{r}))$. 


Next, 
since $\hat \Sigma_t $ in  (\ref{Sigma-def2}) is estimated  as $O(\ep)$, 
it is found from (\ref{NNRF}) that $J_t^{\alpha a}{}^{(0)} = 
\bra \hat J^{\alpha a}\ket_{\lambda_t}^{\rm LG}$.
From the isotropic property, $\bra \hat J^{0a}{}' \ket^{\rm LG}_{\lambda}=0$
and $\bra \hat J^{ab}{}' \ket^{\rm LG}_{\lambda}=\phi\delta^{ab}$,
where $\phi(\bv{r})$ is a scalar field. Then, 
the combination of (\ref{reversible}) with (\ref{sigma-def})
leads to $\beta \partial^a u^a \cdot(\phi-p)=0$.
Since $\phi-p$ is independent of $\bv{u}$, $\phi=p$ is obtained. 
The result was in fact  derived  by the direct calculation \cite{Kirkwood}.  
It is thus concluded that 
$J^{0a}{}^{(0)} = (h+p)u^a $, $J^{ab}{}^{(0)}
= p \delta^{ab}+ \rho u^a u^b$, and 
$J^{4a}=J^{4a}{}^{(0)}=\rho u^a$,
where the relation $\hat J^{\alpha a}{}'$ and $\hat J^{\alpha a}$
has been used \cite{supp}. 
The hydrodynamic equation 
$\partial_t C^\alpha_t+\partial^a J_t^{\alpha a}{}^{(0)}=0$
is the Euler equation. (See Ref. \cite{Yau} for the mathematical
derivation of the Euler equation.)


Furthermore, from (\ref{NNRF}), it is found that $J_t^{\alpha a}{}^{(1)} = 
\bra \hat J^{\alpha a} \hat \Sigma_t^{(1)}\ket_{\lambda_t}^{\rm LG}$,
where $\hat \Sigma_t^{(1)}$ represents the estimation that 
$\partial_s \lambda_s$ in $\hat \Sigma_t$ is given by 
the Euler equation. Then, 
$\hat \Sigma^{(1)}_t$ is calculated as \cite{supp}
\begin{equation}
\hat \Sigma^{(1)}_t(\Gamma)
= \int_0^t ds 
[ (\partial^a \beta_s)\cdot \hat q^{a}(\Gamma_{s-t})
 -(\beta_s \partial^a u^b_s) \cdot \hat \tau^{ab}(\Gamma_{s-t})],
\label{Sigma-def3} 
\end{equation}
where
\begin{eqnarray}
\hat q^a &\equiv&  \hat J^{0a}{}' - \frac{h'+p}{\rho} \hat \pi^a{}',
\label{qdef}\\
\hat \tau^{ab} &\equiv & 
\delta \hat J^{ab}{}' - \left[\pderf{p}{h'}{\rho} \delta \hat h' + 
\pderf{p}{\rho}{h'} \delta \hat \rho \right] \delta^{ab}.
\label{taudef}
\end{eqnarray}
$\hat q^{a}$ and $\hat \tau^{ab}$ are interpreted as 
the  irreducible part of the energy  and momentum fluxes
in the moving frame  with the local velocity,
in which  the contribution from fluctuations of
the density fields are subtracted from the 
energy  and momentum fluxes.
By combining  (\ref{qdef}) and (\ref{taudef}) with
the relation between $\hat J^{\alpha a}{}'$ and $\hat J^{\alpha a}$,
it is obtained that $J^{0a}_t{}^{(1)} = \tau_t^{ab}{}^{(1)}u_t^b+q_t^a{}^{(1)}$
and $J^{ab}_t{}^{(1)} =\tau_t^{ab}{}^{(1)}$. 
Here, it is assumed that the characteristic length and time 
scales of $C^{\alpha}$ are much larger than the correlation 
length and time of $\hat \tau^{ab}{}^{(1)}$  and $\hat q^a{}^{(1)}$. 
This enables us to replace $C^\alpha_s(\bv{r}')$ in 
the space-time integration of 
$\bra \hat q^{ a}{}^{(1)} \hat \Sigma_t^{(1)}\ket_{\lambda_t}^{\rm LG}$
and
$\bra \hat\tau^{ab}{}^{(1)} \hat \Sigma_t^{(1)}\ket_{\lambda_t}^{\rm LG}$
by $C^\alpha_t(\bv{r})$. 
By noting the isotropic property of the system, it is finally
derived that 
$q_t^a{}^{(1)}=\kappa (\partial^a \beta_t)$ and 
\begin{equation}
\tau^{ab}_t{}^{(1)} = -\eta (\partial^a   u^b_t+\partial^b   u^a_t)
- \left(\zeta-\frac{2}{3}\eta \right) \partial^c  u_t^c \delta^{ab},
\end{equation}
with the Green-Kubo formula \cite{Green}
\begin{eqnarray}
\kappa &=& \int_0^t ds \int d^3 \bv{r}'  
\bra  \hat q^{1}(\bv{r}';\Gamma_{s-t})  \hat q^{1}(\bv{r};\Gamma) 
\ket_{\lambda_t}^{\rm LG}, 
\label{kappa}\\
\eta &=& \beta_t \int_0^t ds \int d^3 \bv{r}'  
\bra \hat \tau^{12}(\bv{r}';\Gamma_{s-t}) \hat \tau^{12}(\bv{r};\Gamma) 
\ket_{\lambda_t}^{\rm LG},  
\label{eta} \\
\zeta &=& \beta_t \int_0^t ds \int d^3 \bv{r}'  
\bra \delta \hat p(\bv{r}';\Gamma_{s-t}) \delta \hat p(\bv{r};\Gamma) 
\ket_{\lambda_t}^{\rm LG}, 
\label{zeta}
\end{eqnarray}
where $\delta\hat p \equiv \hat \tau^{aa}{}^{(1)}/3$. 
When $t$ is much larger than the correlation time of $\hat \tau^{ab}{}^{(1)}$ 
and $\hat q^a{}^{(1)}$, the transportation coefficients $\kappa$, $\eta$, and 
$\zeta$ 
depend on $t$ only through the $t$-dependence of  $\beta_t$ and $\mu_t$,
but are independent of $\bv{u}$.
The hydrodynamic equation $\partial_t C^\alpha_t +\partial^a [J_t^{\alpha a}
{}^{(0)}+J_t^{\alpha a}{}^{(1)}]=0$ is the Navier-Stokes equation,
where $J^{4a}_t{}^{(1)} = 0$.
As is understood from the derivation method, the equation is valid
up to $t \simeq O(\ep^{-2})$, which is enough to describe relaxation 
processes to the global equilibrium state. 
The expression (\ref{pt-2}) with $\hat \Sigma_t \simeq O(\epsilon)$ implies
that the distribution $P_t(\Gamma)$ is close to the local equilibrium 
in the relaxation processes.


It should be noted that  
the well-defined nature of the Green-Kubo formulas 
for non-uniform systems, (\ref{kappa}), (\ref{eta}), and (\ref{zeta}),
remain to be studied seriously, in particular, with regard to 
their relevance to the power-law behavior of the time correlation 
functions in the formulas \cite{Pomeau}.
This is also related to the study of fluctuations
of the macroscopic density fields defined as 
the average of $\hat C^\alpha$ over a spherical 
region with radius $\Lambda$, where $\Lambda$ satisfies 
$ \xi_{\rm micro} \ll \Lambda \ll \xi_{\rm macro}$.
Here, on one hand,
because $ \xi_{\rm micro} \ll \Lambda$, it is expected from 
the law of large numbers that $\hat C_{\Lambda}^\alpha(\bv{r};\Gamma)$ 
takes a typical value with respect to the probability density $P_t(\Gamma)$. 
On the other hand, the condition $ \Lambda \ll \xi_{\rm macro}$ leads 
to the result that the typical value of 
$\hat C_{\Lambda}^\alpha(\bv{r};\Gamma)$ is independent of 
the cut-off length $\Lambda$. In this sense, 
$\hat C_{\Lambda}^\alpha(\bv{r};\Gamma)$ 
defines macroscopic density fields without ambiguity.
Since their typical value is expected  to be given by 
$C^\alpha_t(\bv{r})$, the deterministic part of the 
evolution equation for the macroscopic density fields 
is the hydrodynamic equation. To construct a simple
formulation for describing statistical properties  
of $\hat C_{\Lambda}^\alpha(\bv{r};\Gamma)$
is the next natural problem.


\paragraph{Concluding remarks:}


In summary, the Navier-Stokes  equation has been derived from 
Hamiltonian particle systems in the most compact manner. 
The simple method presented in this Letter is expected 
to be useful for the derivation of hydrodynamic equations
in other systems such as relativistic systems \cite{Minami,Kunihiro}, 
an-isotropic molecular systems \cite{LC}, visco-elastic 
systems \cite{Ooshida,Kawai,Fukuma}, dissipative particles
\cite{granular}, and active matter \cite{active}.
The formulation may also be developed so as to describe more 
complicated behavior near boundaries. It is stimulating
to theoretically study a limit of the hydrodynamic 
non-slip boundary conditions \cite{slip}.  


Another important theoretical problem to be solved is to derive the 
hydrodynamic equations for other initial distributions. 
However, it seems quite difficult to study the problem,
and a new concept would be necessary for a breakthrough. In particular,
a role of the local equilibrium distributions may be clarified more.
Regarding this difficulty, neither a proof of the non-decreasing
property of $\cS(C_t)$ nor a counter example is obtained, 
whereas $ \cS^{(0)}(C_t)$ is found to be non-decreasing 
when the Navier-Stokes equation 
is considered. The understanding of $\cS(C)$ beyond $\cS^{(0)}(C)$
may be connected to steady-state thermodynamics
\cite{Oono,Hatano-Sasa,Ruelle,Sasa-Tasaki,KNST,KNST-nl,
Nakagawa,Christian,Jona-Lasinio,Sasa2}.

I hope that the formulation described in this Letter
will be developed further in order to 
discover the relationship between  the macroscopic and microscopic 
descriptions.


The author thanks Y. Yokokura, M. Fukuma, and T. Kunihiro 
for stimulating discussions
on the foundation of  hydrodynamics. The author also thanks
Y. Nakayama, T. Nakamura, Y. Oono,
T. Ooshida, M. Otsuki, K. Saitoh, T. Sasamoto, N. Shiraishi, K. Takeuchi, 
H. Tasaki, and H. Watanabe for useful comments on the manuscript. 
The present study was supported by KAKENHI Nos. 22340109, 25103002,
and by the JSPS Core-to-Core program 
``Non-equilibrium dynamics of soft-matter and information''.


　
\eject
\onecolumngrid

\def\theequation{S\arabic{equation}}
\makeatletter
\@addtoreset{equation}{section}
\makeatother

\setcounter{equation}{0}


\section{Supplemental Material}

\subsection{Microscopic current}

The microscopic current $\hat J^{\alpha a}(\bv{r};\Gamma)$
is determined from the Hamiltonian under consideration.
The explicit expression is given here. Let $x_i^a$, $x^a$, and 
$p_i^a$ be the components of $\bv{r}_i$, $\bv{r}$, 
and $\bv{p}_i$, respectively; and  $F_{ij}$ and $D$ are 
defined as 
\begin{equation}
F_{ij}^a=-\pder{V(|\bv{r}_{ij}|)}{x_i^a},
\end{equation}
and
\begin{equation}
D(\bv{r};\bv{r}_i,\bv{r}_j)
=\int_0^1  d\xi 
\delta(\bv{r}-\bv{r}_i-(\bv{r}_j-\bv{r}_i)\xi).
\end{equation}
The direct calculation of $\partial_t \hat C^\alpha(\bv{r};\Gamma_t)$ 
yields
\begin{eqnarray}
\hat J^{0a} 
&=& \sum_i 
\left[\frac{p_i^2}{2m} +\frac{1}{2}\sum_{j \not =i} V(|\bv{r}_{ij}|) \right]
\frac{p_i^a}{m}\delta(\bv{r}-\bv{r}_i)
+ \frac{1}{2} \sum_{i < j} \frac{p_i^b+p_j^b}{m} F_{ij}^b (x_i^a-x_j^a) 
D(\bv{r};\bv{r}_i,\bv{r}_j), \\
\hat J^{ab} 
&=& \sum_i \frac{p_i^ap_i^b}{m}\delta(\bv{r}-\bv{r}_i(t)) 
+  \sum_{i < j} F_{ij}^a (x_i^b-x_j^b) 
D(\bv{r};\bv{r}_i,\bv{r}_j),
\\
\hat J^{4a}&=&\hat \pi^a. 
\end{eqnarray}
Here, the following identity has been used:
\begin{equation}
\delta(\bv{r}-\bv{r}_i(t))-\delta(\bv{r}-\bv{r}_j(t))
= \pder{}{x^b}(x_j^b-x_i^b) D(\bv{r};\bv{r}_i,\bv{r}_j).
\end{equation}
It can be confirmed that $\hat J^{ab}=\hat J^{ba}$.
From the expressions of $\hat J^{\alpha a}$, the microscopic 
current in the moving frame $\hat J^{\alpha a}{}'$ is
obtained as
\begin{eqnarray}
\hat J^{4a}{}'&=& \hat \pi^a-\hat \rho u^a, \\
\hat J^{ab}{}'
&=& \hat J^{ab}+ \hat \rho u^a u^b -\hat \pi^a u^b-\hat \pi^b u^a, \\
\hat J^{0a}{}' 
&=& \hat J^{0a} - \hat J^{ab}{}' u^b -\hat \pi^a{}' \frac{u^2}{2}-\hat h u^a
+\hat \Delta^a,
\end{eqnarray} 
where 
\begin{equation}
\hat \Delta^a(\bv{r};\Gamma)=
\sum_{i <j} (u^b(\bv{r})-u^b(\bv{r}_i))
F_{ij}^b (x_i^a-x_j^a) 
D(\bv{r};\bv{r}_i,\bv{r}_j).
\end{equation}


\subsection{Derivation of (18) and (19)}

In this section, (18) and (19) in the main text are derived.   

\paragraph{Preliminaries:}

From  the definitions  $\nu\equiv \beta \mu$ and 
$\psi \equiv \beta p=s_{\rm th}-\beta h'+\nu \rho$,
the relation 
\begin{equation}
d \psi= - h' d \beta  + \rho d \nu
\label{thermo}
\end{equation}
is obtained. This leads to 
\begin{eqnarray}
\pderf{p}{\beta}{\nu} &=& -\frac{h'+p}{\beta}, 
\label{pbeta}\\
\pderf{p}{\nu}{\beta} &=& \frac{\rho}{\beta}.
\label{pnu}
\end{eqnarray}
The Euler equation can be written as 
\begin{eqnarray}
(\partial_t +u^a\partial^a ) \rho &=&  -\rho (\partial^a u^a), 
\label{rho-eq}\\
\rho (\partial_t +u^a\partial^a ) u^b &=&  -\partial^b p, 
\label{u-eq} \\
(\partial_t +u^a\partial^a ) h' &=&  -(h'+p) (\partial^a u^a).
\label{h-eq}
\end{eqnarray}
Let $f(h', \rho)$ be any function of $h'$ and $\rho$.
By using (\ref{pbeta}), (\ref{pnu}), 
(\ref{rho-eq}), and (\ref{h-eq}), the time evolution equation of 
$f$ is derived as
\begin{equation}
(\partial_t +u^a\partial^a ) f = g \beta (\partial^a u^a)
\end{equation}
with
\begin{equation}
g=
\pderf{f}{h'}{\rho}\pderf{p}{\beta}{\nu}-
\pderf{f}{\rho}{h'}\pderf{p}{\nu}{\beta}.
\end{equation}
By setting $f=\beta$ and $f=\nu$, the following
equations are obtained, respectively:
\begin{eqnarray}
(\partial_t +u^a\partial^a ) \beta 
&=& \beta \pderf{p}{h'}{\rho}(\partial^a u^a), 
\label{beta-eq}\\
(\partial_t +u^a\partial^a ) \nu 
&=& -\beta \pderf{p}{\rho}{h'}(\partial^a u^a), 
\label{nu-eq}
\end{eqnarray}
where  the Maxwell relation
\begin{equation}
\pderf{\beta}{\rho}{h'}=-\pderf{\nu}{h'}{\rho}
\end{equation}
was used in the derivation.
Hereafter, $O(\ep^3)$ terms are neglected in the calculations. 

\paragraph{Derivation of (18):}
By substituting $J^{0a}=J^{0a}{}'+J^{ab}{}'u^b +h u^a$
and $J^{ab}=J^{ab}{}'+\rho u^a u^b$ into 
$\partial^a \lambda^\alpha \cdot J^{\alpha a}$, 
\begin{equation}
\partial^a \lambda^\alpha \cdot J^{\alpha a}
= (\partial^a \beta) \cdot J^{0a}{}'-(\beta \partial^a u^b)\cdot J^{ab}{}'
-u^a \cdot (\partial^a \psi),
\end{equation}
where (\ref{thermo}) was used. By recalling $\psi=\beta p$,
(18) in the text has been obtained. 

\paragraph{Derivation of (19):}
The estimation of $\lambda$ in terms of $\beta$ and $\nu$ yields
\begin{eqnarray}
(\partial^a \lambda) \cdot\hat J^{\alpha a} 
&=&
 (\partial^a \beta) \cdot
( \hat J^{0a}-u^b\hat J^{ba}+\frac{u^2}{2}\hat \pi^a) 
 -(\partial^a \nu) \cdot \hat \pi^a 
+(\beta \partial^a u^b) \cdot (\hat \pi^au^b-\hat J^{ba}),
\label{eq1}
\end{eqnarray}
and 
\begin{eqnarray}
(\partial_t \lambda) \cdot \hat C^{\alpha } 
&=& 
 (\partial_t \beta) \cdot
\left(\hat h-u^a \hat \pi^{a}+\frac{u^2}{2}\hat \rho \right)  
 - (\partial_t \nu) \cdot\hat \rho 
-(\beta \partial_t u^a) \cdot \hat \pi^a{}'.
\label{eq2}
\end{eqnarray}
Further, (\ref{u-eq}) with  (\ref{pbeta}) and (\ref{pnu}) leads to
\begin{eqnarray}
-(\beta \partial_t u^a) \cdot \hat \pi^{a}{}'  
&= &  - \frac{h'+p}{\rho}(\partial^a \beta) \cdot \hat \pi^a{}' 
 +(\partial^a \nu) \cdot \hat \pi^a{}'
 +(\beta u^b \partial^b u^a) \cdot \hat \pi^a{}'.
\label{eq3}
\end{eqnarray}
By summing (\ref{eq1}), (\ref{eq2}), and (\ref{eq3}),
using (\ref{beta-eq}) and (\ref{nu-eq}), and subtracting
the expectation with respect to $P_{\rm LG}$, (19) with
(20) and (21) in the main text has been obtained.


\begin{thebibliography}{10}


\bibitem{Landau}
L. D. Landau and E. M. Lifshitz, {\it Fluid Mechanics}, 
(Pergamon Press, Oxford, 1959).


\bibitem{Komatsu}
T. S. Komatsu, S. Matsumoto, T. Shimada, and N. Ito,
``A glimpse of fluid turbulence from the molecular scale'',
Int. J. Mod. Phys. C, to be published 2014, DOI: 10.1142/S012918311450034X.



\bibitem{Grad}
H. Grad, Phys. Fluids {\bf 6}, 147 (1963).



\bibitem{Kirkwood}
J. H. Irving and J. G. Kirkwood,
J. Chem. Phys. {\bf 18}, 817 (1950).

\bibitem{Morray}
C. B. Morray, Commun. Pure and Appl. Math. {\bf 8}, 279 (1955).



\bibitem{Mori}
H. Mori,
Phys. Rev. {\bf 112}, 1829 (1958).

\bibitem{Mclennan}
J. A. Mclennan, Phys. Fluids {\bf 3}, 493 (1960);
{\it Introduction to Non-equilibrium Statistical Mechanics}
(Prentice-Hall, 1988).

\bibitem{Zubarev}
D. N. Zubarev,
{\it Nonequilibrium Statistical Thermodynamics}
(Consultants Bureau, New York, 1974).

\bibitem{KG}
K. Kawasaki and J. D. Gunton,
Phys. Rev. A {\bf 8}, 2048 (1973). 


\bibitem{Maes2}
C. Maes and K. Neto\v{c}n\'{y}, 
J. Math. Phys. {\bf 51}, 015219 (2010).



\bibitem{Spohn-book}
H. Spohn, {\it Large Scale Dynamics of Interacting Particles}, 
(Texts and Monographs in Physics, Springer-Verlag, Heidelberg, 1991). 



\bibitem{Esposito}
R. Esposito and R. Marra,
J. Stat. Phys. {\bf 74},  981 (1994).


\bibitem{Lenard}
A. Lenard, J. Stat. Phys. {\bf 19}, 575 (1978).

\bibitem{Tasaki}
H. Tasaki, Statistical mechanical derivation 
of the second law of thermodynamics,
preprint, http://arxiv.org/abs/cond-mat/0009206

\bibitem{Jarzynski}
{C. Jarzynski}, { Phys. Rev. Lett.} \textbf{78}, 2690 (1997).


%


\bibitem{Evans}
D. J. Evans,  E. G. D. Cohen,  and G. P. Morriss,
{Phys. Rev. Lett.} 
\textbf{71},  2401 (1993).

\bibitem{Gallavotti}
G. Gallavotti and E. G. D. Cohen,
{ Phys. Rev. Lett.}
\textbf{74},   2694 (1995).

\bibitem{Kurchan}
J. Kurchan,  
{J. Phys. A: Math. Gen.}
\textbf{31},   3719 (1998).

\bibitem{Maes}
C. Maes, {J. Stat. Phys.}  
\textbf{95},  367 (1999).

\bibitem{LS}
J. L. Lebowitz and  H. Spohn, {J. Stat. Phys.}  
\textbf{95}, 333 (1999). 


\bibitem{Crooks}
{G. E. Crooks},  Phys. Rev. E 
\textbf{61}, {2361} (2000).

\bibitem{d-FT}
C. Jarzynski, J. Stat. Phys. {\bf 98}, 77 (2000).

\bibitem{Seifert}
{U. Seifert}, 
{Phys. Rev. Lett.} \textbf{95}, {040602} {(2005)}.




\bibitem{supp}
See Supplemental Material at [URL will be
inserted by publisher] .



\bibitem{Yau}
S. Olla, S. R. S. Varadhan, and H. T. Yau,
Commun. Math. Phys. {\bf 155} 523 (1993).



\bibitem{Green}
M. S. Green,
J. Chem. Phys. {\bf 22}, 398 (1954).

\bibitem{Pomeau}
Y. Pomeau and P. Resibois, 
Phys. Rep. {\bf 19}, 63 (1975).


\bibitem{Minami}
Y. Minami and Y. Hidaka,
Phys. Rev. E {\bf 87}, 023007 (2013).

\bibitem{Kunihiro}
K. Tsumura  and T. Kunihiro,
Phys. Rev. E {\bf 87}, 053008 (2013).



\bibitem{LC}
P. C. Martin, O. Parodi, and P. S.  Pershan, 
Phys. Rev. A {\bf 6}, 2401 (1972). 


\bibitem{Ooshida}
T. Ooshida and K. Sekimoto,
Phys. Rev. Lett. {\bf 95}, 108301 (2005).

\bibitem{Kawai}
T. Azeyanagi, M. Fukuma, H. Kawai, and K. Yoshida, 
Physics Letters B {\bf 681}, 290 (2009).

\bibitem{Fukuma}
M. Fukuma and Y. Sakatani, 
Phys. Rev. E {\bf 84}, 026316 (2011).


\bibitem{granular}
J. J.  Brey, J. W.  Dufty, C. S. Kim, and A. Santos,
Phys. Rev. E  {\bf 58}, 4638 (1998). 


\bibitem{active}
M. C. Marchetti, J. F. Joanny, S. Ramaswamy, T. B. Liverpool, J. Prost, 
M. Rao, and R. A. Simha,
Rev. Mod. Phys. {\bf 85}, 1143 (2013). 

\bibitem{slip}
Y. Zhu and S. Granick, 
Phys. Rev. Lett. {\bf 88}, 106102 (2002). 



\bibitem{Oono}
{Y. Oono and M. Paniconi},
{Prog. Theor. Phys. Suppl.}
\textbf{130}, 29 (1998).

\bibitem{Hatano-Sasa}
{T. Hatano and S.-i. Sasa},
{Phys. Rev. Lett.}
\textbf{86}, 3463 
{(2001)}.

\bibitem{Ruelle}
D. Ruelle,
Proc. Natl. Acad. Sci. U.S.A.  {\bf 100}, 3054 (2003).

\bibitem{Sasa-Tasaki}
S.-i. Sasa and H. Tasaki,
J. Stat. Phys. {\bf 125}, 125 (2006).


\bibitem{KNST}
T. S. Komatsu, N. Nakagawa, S.-i. Sasa, and H. Tasaki,
Phys. Rev. Lett., {\bf 100}, 230602 (2008). 

\bibitem{KNST-nl}
T. S. Komatsu, N. Nakagawa, S.-i. Sasa, and H. Tasaki,
J. Stat. Phys. {\bf 142}, 127 (2011).

\bibitem{Nakagawa}
N. Nakagawa,
Phys. Rev. E {\bf 85}, 051115 (2012).

\bibitem{Christian}
{C. Maes and K. Neto\v{c}n\'{y}}, 
arXiv:1206.3423.

\bibitem{Jona-Lasinio}
E. Bertini, D. Gabrielli, G. Jona-Lasinio, and C. Landim,
Phys. Rev. Lett., {\bf 110}, 020601 (2013).

\bibitem{Sasa2}
S.-i. Sasa,  J. Stat. Mech., P01004 (2014).








\end{thebibliography}
\end{document}